\documentclass[a4paper,20pt]{article}
    \usepackage[T1]{fontenc}
    \usepackage{graphicx}
    \usepackage{epsfig}
    \usepackage{amsmath}
    \usepackage{amsfonts}
    \renewcommand{\abstract}{}
\begin{document}
\makeatletter
\renewcommand{\@oddhead}{\textit{Advances in Astronomy and Space Physics} \hfil \textit{R. \v{S}m\'{i}da for the Pierre Auger Collaboration}}
\renewcommand{\@evenfoot}{\hfil \thepage \hfil}
\renewcommand{\@oddfoot}{\hfil \thepage \hfil}
\fontsize{11}{11} \selectfont

\title{Results from the Pierre Auger Observatory}
\author{
\textsl{R. \v{S}m\'{i}da$^{1}$ for the Pierre Auger Collaboration$^{2}$}
}
\date{}
\maketitle
\begin{center} 
{\small $^{1}$Karlsruhe Institute of Technology (KIT), Hermann-von-Helmholtz-Platz 1, 76344 Eggenstein-Leopoldshafen, Germany\\
$^{2}$Observatorio Pierre Auger, Av. San Mart\'{i}n Norte 304, 5613 Malarg\"{u}e, Argentina\\
Full author list: http://www.auger.org/archive/authors\_2011\_10.html\\
Email: auger\_spokespersons@fnal.gov}
\end{center}

\begin{abstract}
The Pierre Auger Observatory is the largest observatory of high-energy cosmic rays.
It is located in Argentina and has been taking data since January 2004. Extensive
air showers initiated by cosmic rays are measured by the hybrid
detector, which combines the sampling of particle density at ground by water-Cherenkov
tanks and the measurement of atmospheric fluorescence light by telescopes. New detection
techniques, like radio and microwave measurement, are also being tested. Results
regarding the energy spectrum, mass composition and arrival directions of cosmic
rays are presented here.
\end{abstract}

\section*{Introduction}
The goal of the Pierre Auger Observatory~\cite{pao} is the measurement of ultra-high energy cosmic rays (UHECRs), i.e. cosmic rays
with energies above $10^{18}$\,eV. The origin and properties of UHECRs are not fully understood yet. These particles have been observed
for more than half a century and special attention has been paid to the most energetic ones.

Unlike optical or radio photons, the cosmic rays are charged nuclei and are deflected by magnetic fields during their propagation
through interstellar and intergalactic space. Our knowledge of the galactic and extragalactic magnetic fields is limited. Since the deflections
caused by the magnetic fields are inversely proportional to the particle energy, we might take advantage of using the most energetic
measured cosmic rays for backtracking to a source location.

The investigation of UHECRs is difficult and requires a large effort. First of all, the flux\footnote{
The flux is defined as the number of cosmic rays arriving on a unit area from unit solid angle per unit of time.}
of UHECRs is very low. It falls steeply with energy, decreasing by almost three orders
of magnitude per decade of energy. Only one particle per km$^2$ per year arrives at the Earth above an energy of 10$^{18}$\,eV
and the flux reduces to less than one particle per km$^2$ per century at energies above 10$^{20}$\,eV. It is clear that
a huge detector is necessary for the successful measurement of UHECRs.

Another difficulty in the study of UHECRs is their interaction in the atmosphere. The direct detection of primary particles of cosmic rays
is possible only at altitudes higher than $\sim$20\,km. Balloon or space borne experiments are too limited in their collection area to measure
the low flux of UHECRs and cannot be used. In the 1930s French physicist Pierre Auger~\cite{auger} found
that the primary cosmic ray particle interacts with the atmosphere and many secondary particles are produced
in the first and subsequent interactions. Such a cascade of secondary particles is called an extensive air shower (EAS).
An extensive air shower develops in the atmosphere while the energies of the secondary particles are sufficient to produce new particles.
The number of secondary particles in the shower maximum can be as high as a few billion, but only a small fraction of the secondary
particles arrive at ground level.

Two methods have been established for the measurement of EAS. The first one samples the particle
density at the ground with an array of detectors measuring in coincidence. The other method is a measurement of ultraviolet (UV)
nitrogen fluorescence light emitted along the track of EAS.
The properties of the primary cosmic ray particle are reconstructed from the measured data. A detector using both, a surface and
a fluorescence detector, is called a hybrid detector. Hybrid detection gives more accurate results in comparison with the results
obtained by individual detection techniques.

More details about the measurement and properties of UHECRs can be found in review articles \cite{nagano,bluemer} and
references therein.

\section*{Pierre Auger Observatory}

\begin{figure}[!h]
\centering
\begin{minipage}[t]{.58\linewidth}
\centering
\epsfig{file = 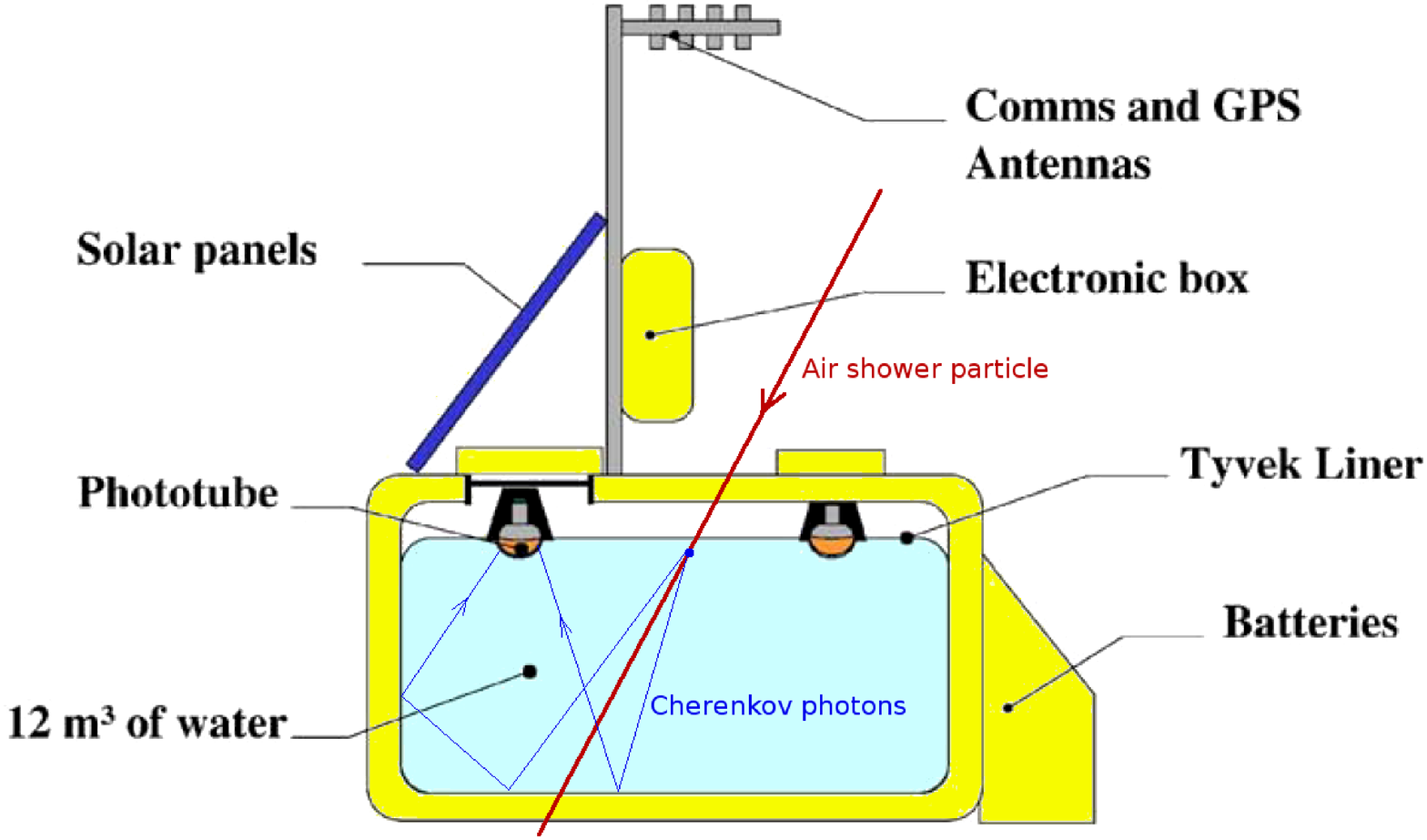,width = .85\linewidth}
\caption{Water-Cherenkov detector with its components. A secondary particle passing through the detector is indicated together
with emitted Cherenkov photons.}\label{fig-sd}
\end{minipage}
\hfill
\begin{minipage}[t]{.4\linewidth}
\centering
\epsfig{file = 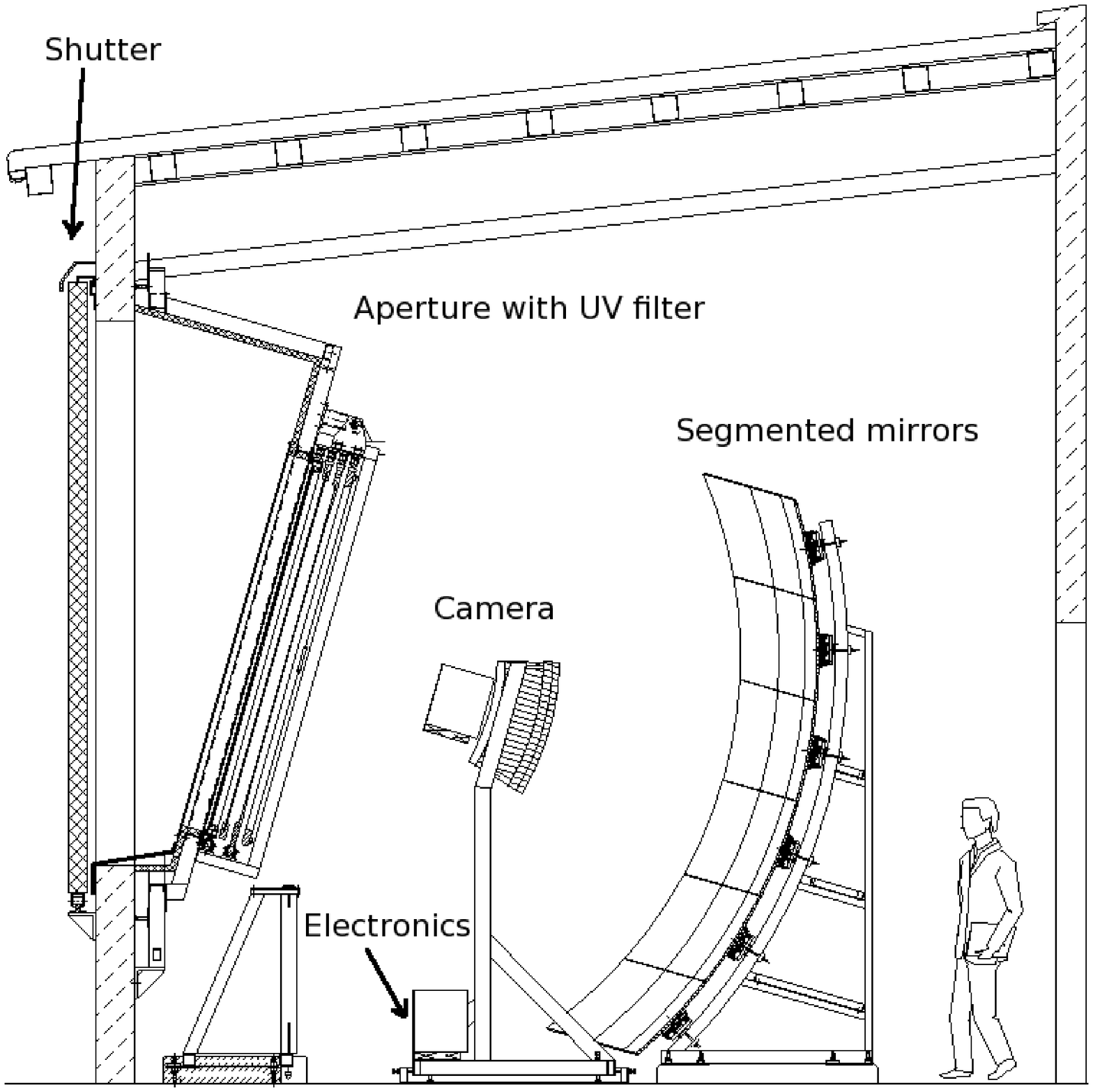,width = .85\linewidth}
\caption{Drawing of a fluorescence telescope. Light passes through the aperture and is reflected by
the mirror to the camera.}\label{fig-fd}
\end{minipage}
\end{figure}

More than 1600 water-Cherenkov surface detectors cover a flat semi-desert area of more than 3000\,km$^2$ near
the city of Malarg\"{u}e, Argentina (69.3$^{\circ}$\,W, 35.3$^{\circ}$\,S, 1400\,m a.s.l.) on a triangular grid with 1.5~km spacing.
Three large photomultipliers measure
Cherenkov photons radiated by relativistic particles passing through purified water in each detector. Each detector is equipped
with necessary electronics, battery, solar panel, GPS and radio comunication antenna (see Fig.~\ref{fig-sd}).

The advantages of the surface detector are its autonomous operation, uptime of almost 100\% and its well defined aperture~\cite{sdpaper}.
On the other hand the energy reconstruction depends on hadronic interaction models, which can only be validated up to energies available
in man-made accelerators. Any usage of hadronic interaction models is necessarily based on theoretical extrapolations from lower
to higher energies. However, using the measurements of the fluorescence detector, the absolute energy scale can be estimated almost
entirely based on measured data.

The fluorescence detector consists of four sites, each with six fluorescence telescopes, located at the boundary of the surface detector array.
The fluorescence telescopes view the atmosphere above the array during clear, almost moonless, nights~\cite{fdpaper}.
The uptime of the fluorescence detector is about 13\%.
Each fluorescence telescope consists of a spherical segmented mirror, a camera with a matrix of 440 photomultipliers and an aperture with
a UV bandpass filter and corrector ring (see Fig.~\ref{fig-fd}).

One of the main advantages of the fluorescence detector is the calorimetric measurement of cosmic ray energies. The charged secondary
particles of an extensive air
shower excite atmospheric molecular nitrogen, which then emits photons isotropically in the UV range (i.e. into several spectral bands
between 300 and 420\,nm). Because the emitted intensity is proportional to the energy deposited by the shower along its path, the energy
reconstruction is independent of hadronic interaction models, except for only a few percent correction for the invisible component
due to muons and neutrinos.
Another advantage is the observation of the longitudinal profile, i.e. the number of secondary particles of EAS as a function of
atmospheric depth. The position of the shower maximum is sensitive to the type of the primary particle. Therefore the chemical
composition of UHECRs can be studied with the fluorescence detector. Knowledge of the chemical composition is crucial for the study
of cosmic ray acceleration and propagation.

Because the fluorescence emission and also light scattering and attenuation depends on the atmospheric conditions between the shower
and the telescope, a large array of atmospheric monitors is operated at the Pierre Auger Observatory~\cite{atmon}. The data
are also used to prevent cloud-obscured data from distorting estimates of the shower energies, shower maxima, and the detector aperture.
Moreover, the sensitivity of the fluorescence detector is regularly monitored using different light sources.

\section*{Cosmic ray energy spectrum}

\begin{figure}[!h]
\centering
\begin{minipage}[t]{.4\linewidth}
\centering
\epsfig{file = 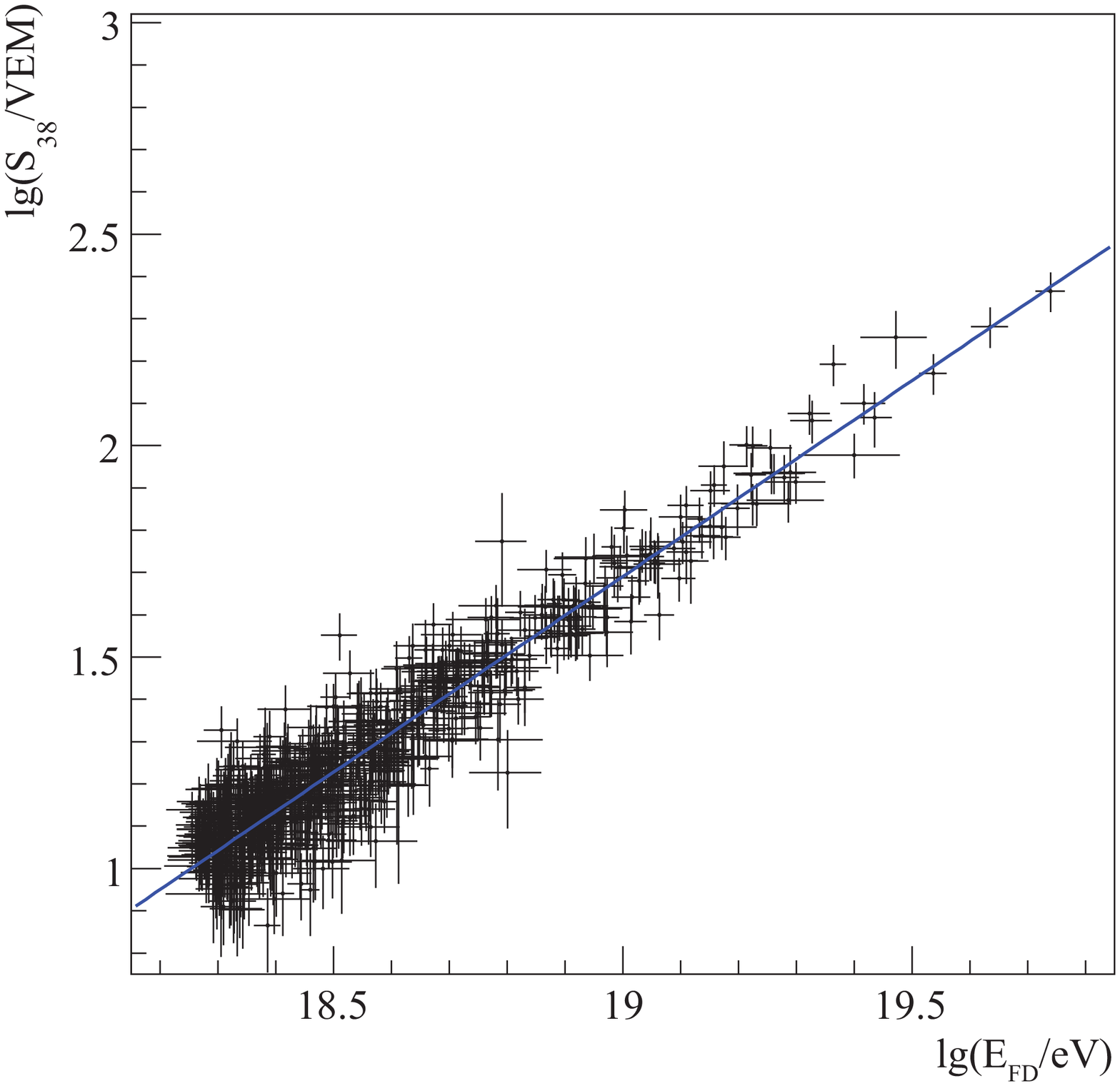,width = .85\linewidth}
\caption{Correlation between the energy estimator of the surface detector 
($S_{38}$) and the energy measured by the fluorescence
detector ($E_{FD}$).
}\label{fig-s38}
\end{minipage}
\hfill
\begin{minipage}[t]{.58\linewidth}
\centering
\epsfig{file = 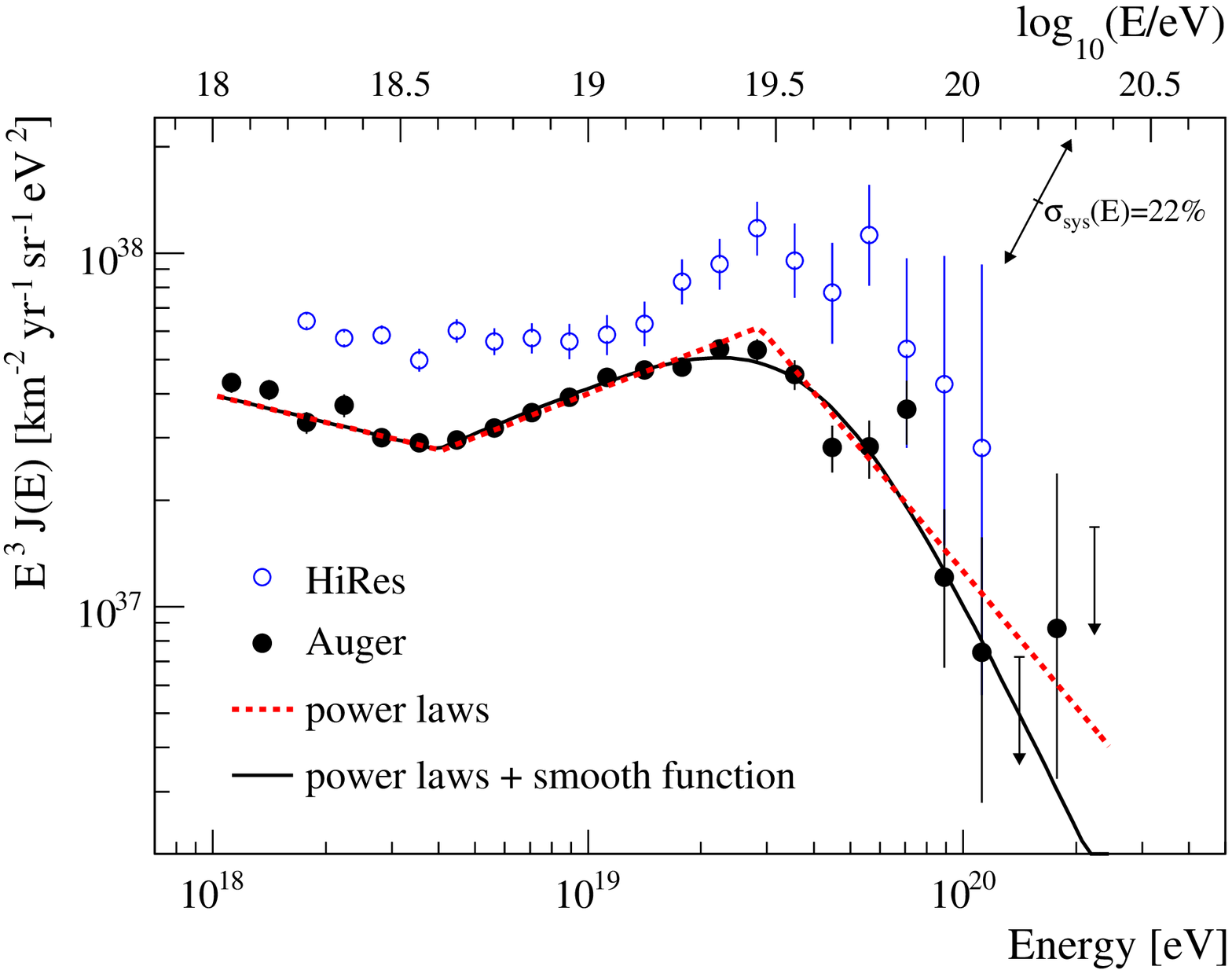,width = .85\linewidth}
\caption{Scaled energy spectrum derived by the Auger experiment and compared with the data from another detector. The systematic
uncertainty of the energy scale of 22\% is indicated by arrows.
}\label{fig-spec}
\end{minipage}
\end{figure}

For the calculation of the cosmic ray flux, a knowledge of the detector exposure together with the precise
energy reconstruction is necessary. The surface detector has a well defined aperture above energies of $3\times10^{18}$\,eV
and, moreover, its exposure does not depend on weather conditions.
The energy estimator of the surface detector is calibrated by the energy given by the fluorescence detector for a subset of EAS
simultaneously detected by both detectors~\cite{spec08,spec10,hybrid}. The correlation between the cosmic ray energy
measured by the fluorescence detector and the energy estimator of the surface detector is shown in Fig.~\ref{fig-s38}.

The combined energy spectrum calculated from the surface and hybrid data is shown in Fig.~\ref{fig-spec}. While the fluorescence
detector provides data below an energy of $3\times10^{18}$\,eV, the surface detector has sufficient statistics even for energies
above a few $10^{19}$\,eV because of its higher uptime. Both spectra show agreement for the intermediate energy range.

The flux of UHECRs follows a power law, with two changes of the spectral index. The flattening of the spectrum
takes place at an energy of $4\times10^{18}$\,eV and this kink is called the ankle. The ankle might indicate the transition
from galactic to extragalactic cosmic rays. A similar feature in the cosmic ray spectrum could also result from the propagation
of protons from extragalactic sources, placing the transition from galactic to extragalactic cosmic rays at a much
lower energy. Various models predict measurable differences not only for the energy spectrum, but also for the chemical composition
(see e.g.~\cite{bluemer}).

A significant suppression of the flux of UHECRs is observed above $4\times10^{19}$\,eV. The suppression is similar
to the prediction of the Greisen-Zatsepin-Kuzmin (GZK) mechanism~\cite{greisen,zatsepin}, but it could also be related
to a change of the injection spectrum in the sources. The GZK mechanism is an interaction of cosmic ray protons above
the GZK energy of $\sim4\times10^{19}$\,eV with the photons of the cosmic microwave background. The energy loss of the interacting
cosmic ray continues until the particle energy falls below the GZK energy. As a result, cosmic rays above the GZK energy cannot
travel over distances larger than $\sim100$\,Mpc without significant energy losses\footnote{Ultra-high energy nuclei lose energy
in photodisintegration on all photon fields.}. This mechanism excludes far distant
sources from making a significant contribution into the UHECR flux above the GZK energy.

The shape of the cosmic ray spectrum differs for proposed models of the origin of UHECRs. It is affected not only by the properties
of the sources but also by propagation processes.

\section*{Cosmic ray composition}

The development of an extensive air shower depends on the type of the primary particle. The longitudinal profile
measured by the fluorescence telescopes and also the lateral distribution of secondary particles sampled by the surface
detector provide information about the properties of the primary particle. It has been found, that the contribution
of neutral particles, which could be interesting for finding the UHECRs sources, to the flux of UHECRs is very low.
The upper limits on the fluxes of known stable neutral particles, photons~\cite{photon} and tau neutrinos~\cite{neutrino},
are so low, that several exotic models of the origin of UHECRs (e.g. the decay of superheavy particles) have been excluded.

The flux of UHECRs is mainly composed of atomic nuclei. The most precise measurement of the chemical composition is obtained
by the observation of the depth of the shower maximum as well as its fluctuations~\cite{comp}.
The EAS produced by lighter primaries (e.g. protons) propagate deeper into the atmosphere than heavier nuclei (e.g. iron nuclei)
and show larger fluctuations. The identification of the primary particle is not possible for a single event, because of the random nature of EAS.

\begin{figure}[!h]
\centering
\epsfig{file = 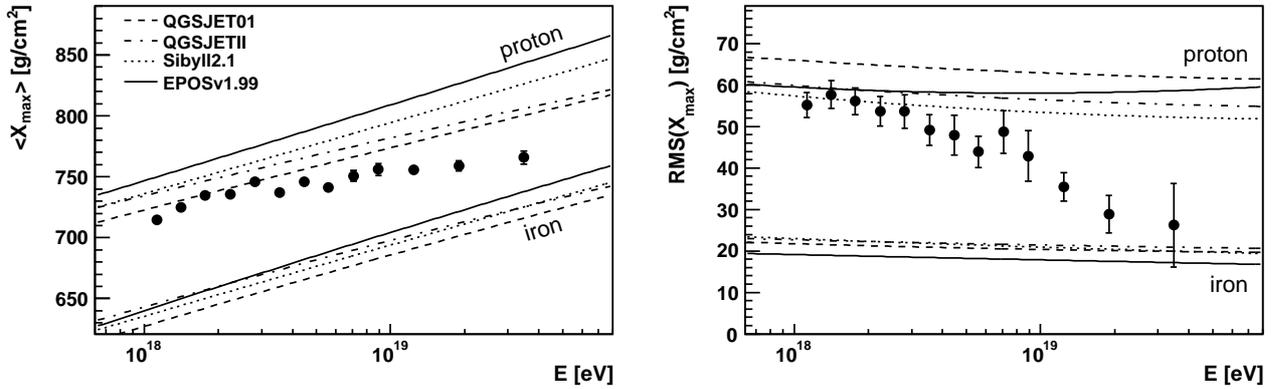,width = .95\linewidth}
\caption{Average depth of shower maximum (left) and shower-to-shower fluctuations (right) as a function of energy
together with air shower simulations using different hadronic interaction models.
}\label{fig-comp}
\end{figure}

The average depth of the shower maximum is found to change with energy as do shower-to-shower fluctuations.
Protons or other light nuclei would be preferable for the study of UHECR sources, because they would be less affected by the magnetic
fields during their propagation from the sources. Unfortunately,
the results of both methods suggest heavier composition at the highest studied energies (see Fig.~\ref{fig-comp}).
It is worth mentioning, that the composition above the energy of $4\times10^{19}$\,eV is not clear yet because of a lack
of statistics.

\section*{Cosmic ray arrival directions}

The uncertainty of reconstructed arrival directions of UHECR by the Pierre Auger Observatory is less than 1.5$^{\circ}$
for events triggering 4 surface detector stations
and better than 1.0$^{\circ}$ for events with six and more stations (i.e. for higher energies). The deflection
of cosmic rays propagating in known magnetic fields is comparable with the angular resolution of the detector even for protons at
the highest energies. The deflection is more than one order of magnitude higher for iron nuclei.

As mentioned above, the sources of UHECRs above energies of $4\times10^{19}$\,eV must lie
closer than $\sim100$\,Mpc. The number of possible astronomical sources (candidate sites are active galactic nuclei (AGN),
radiogalaxies, clusters of galaxies, etc.) within the GZK horizon is limited. These close-by objects are clustered
and we might expect anisotropic arrivals of UHECRs even for a mixed hadronic composition.

\begin{figure}[!h]
\centering
\begin{minipage}[t]{.57\linewidth}
\centering
\epsfig{file = 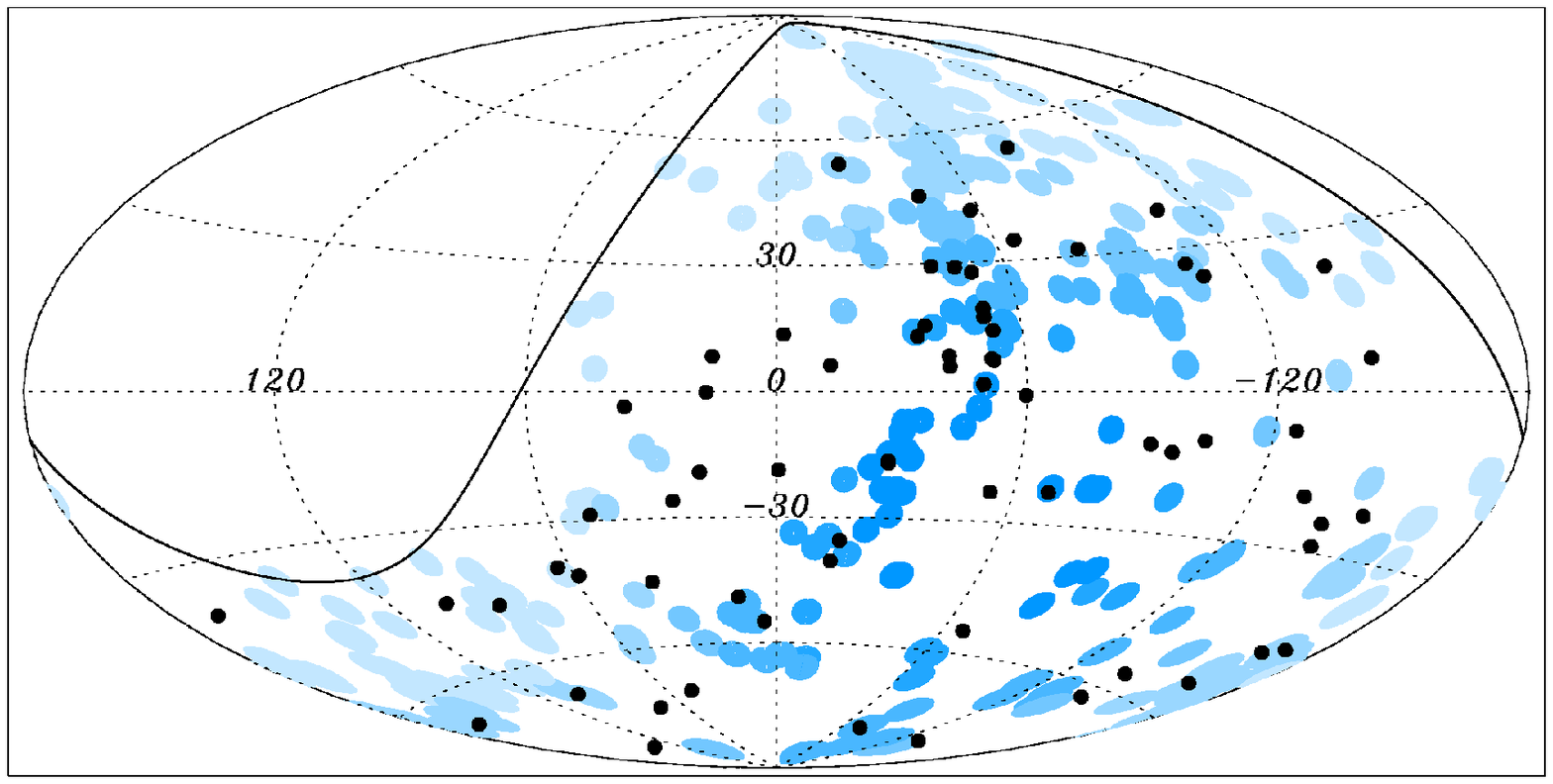,width = .85\linewidth}
\caption{
The arrival directions of 69 events measured by the Pierre Auger Observatory above $5.5\times10^{19}$\,eV (black dots)
together with the positions of nearby AGN (blue points, where darker shade indicates larger relative exposure).
Galactic coordinates are used in the map.
}\label{fig-agn}
\end{minipage}
\hfill
\begin{minipage}[t]{.41\linewidth}
\centering
\epsfig{file=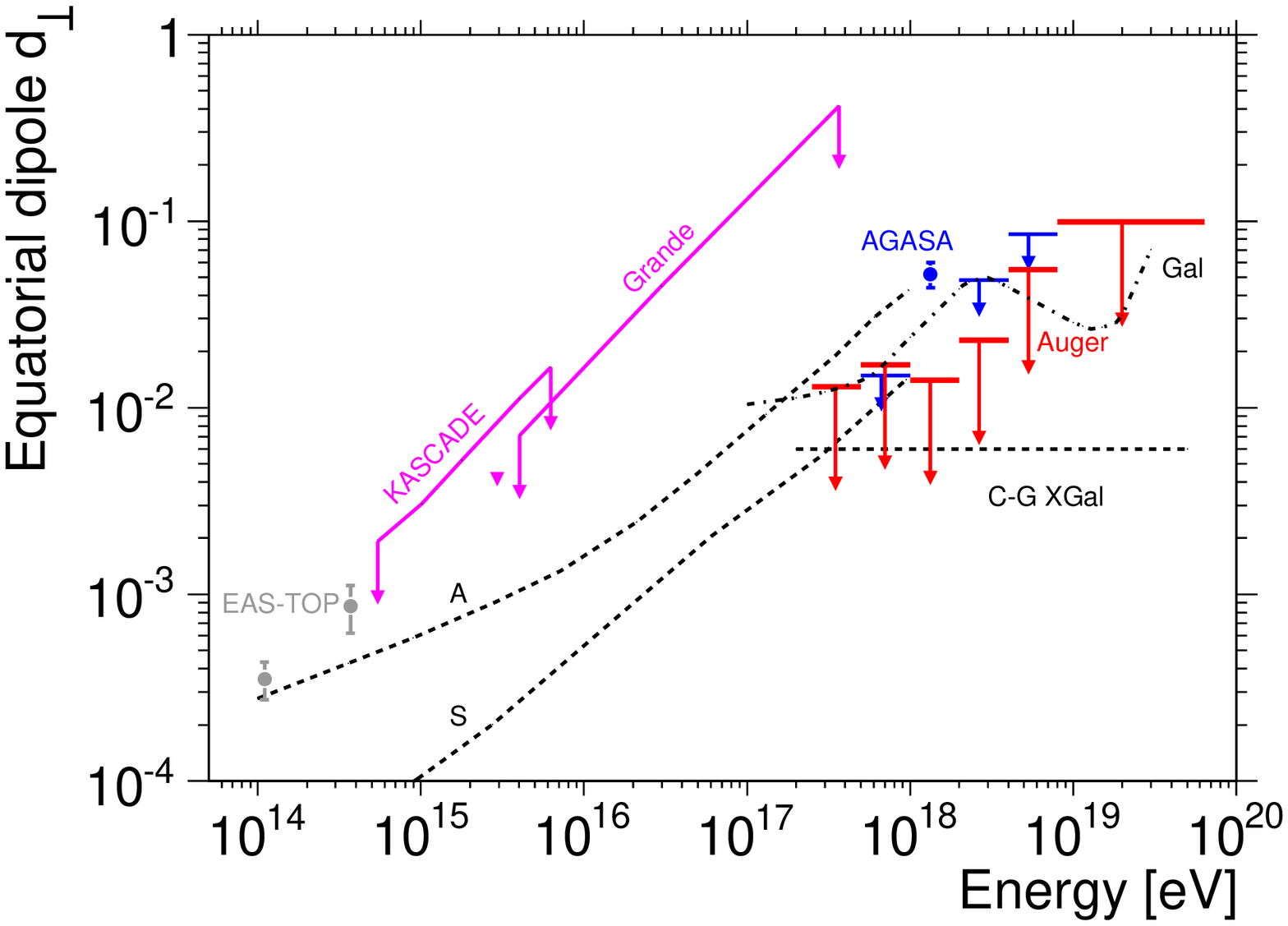,width = .85\linewidth}
\caption{
Upper limits on a dipolar-type anisotropy given by different experiments compared with calculated predictions
(for more detail see~\cite{dipole}).
}\label{fig-dipole}
\end{minipage}
\end{figure}

An anisotropy has been observed by the Pierre Auger Observatory for events above $\sim5.5\times10^{19}$\,eV.
The sky positions of measured arrival directions are preferably closer than $\sim3^{\circ}$ from the positions of AGN
with distances less than 75\,Mpc~\cite{agn07}. The latest results show that the fraction of the highest energetic
cosmic rays correlating with nearby AGN directions is 38\%, while 21\% is expected for isotropic flux, see
Fig.~\ref{fig-agn}. The list of events has been published in~\cite{agn10}. The largest excess has been
measured around the position of our closest radiogalaxy Centaurus~A.

No other excess has been found in the data collected by the Pierre Auger Observatory. The direction towards the Galactic
Centre shows no excess of arrival directions~\cite{galcentre}, nor is there any positive signal found in large-scale anisotropy
studies~\cite{dipole}. The upper limit for the latter result is shown in Fig.~\ref{fig-dipole}.

\section*{Enhancements of the observatory}

\begin{figure}[!h]
\centering
\begin{minipage}[t]{.54\linewidth}
\centering
\epsfig{file=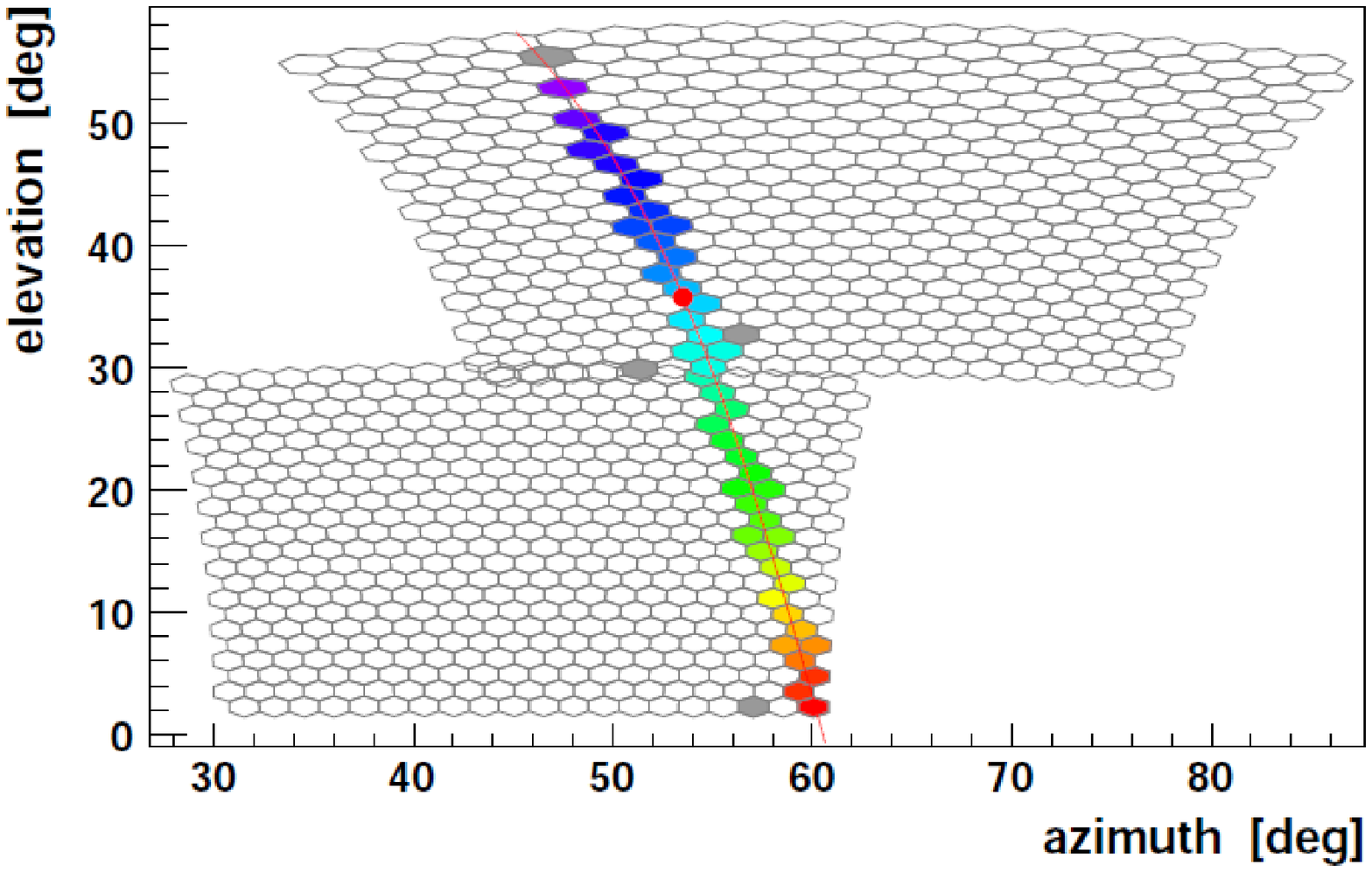,width = .85\linewidth}
\caption{
Trace of a low energetic event measured by two cameras. The upper one belongs to the elevated HEAT telescope, the lower one
is a standard fluorescence telescope. Colors indicate the time evolution of the measured signal.
}\label{fig-heat}
\end{minipage}
\hfill
\begin{minipage}[t]{.44\linewidth}
\centering
\epsfig{file=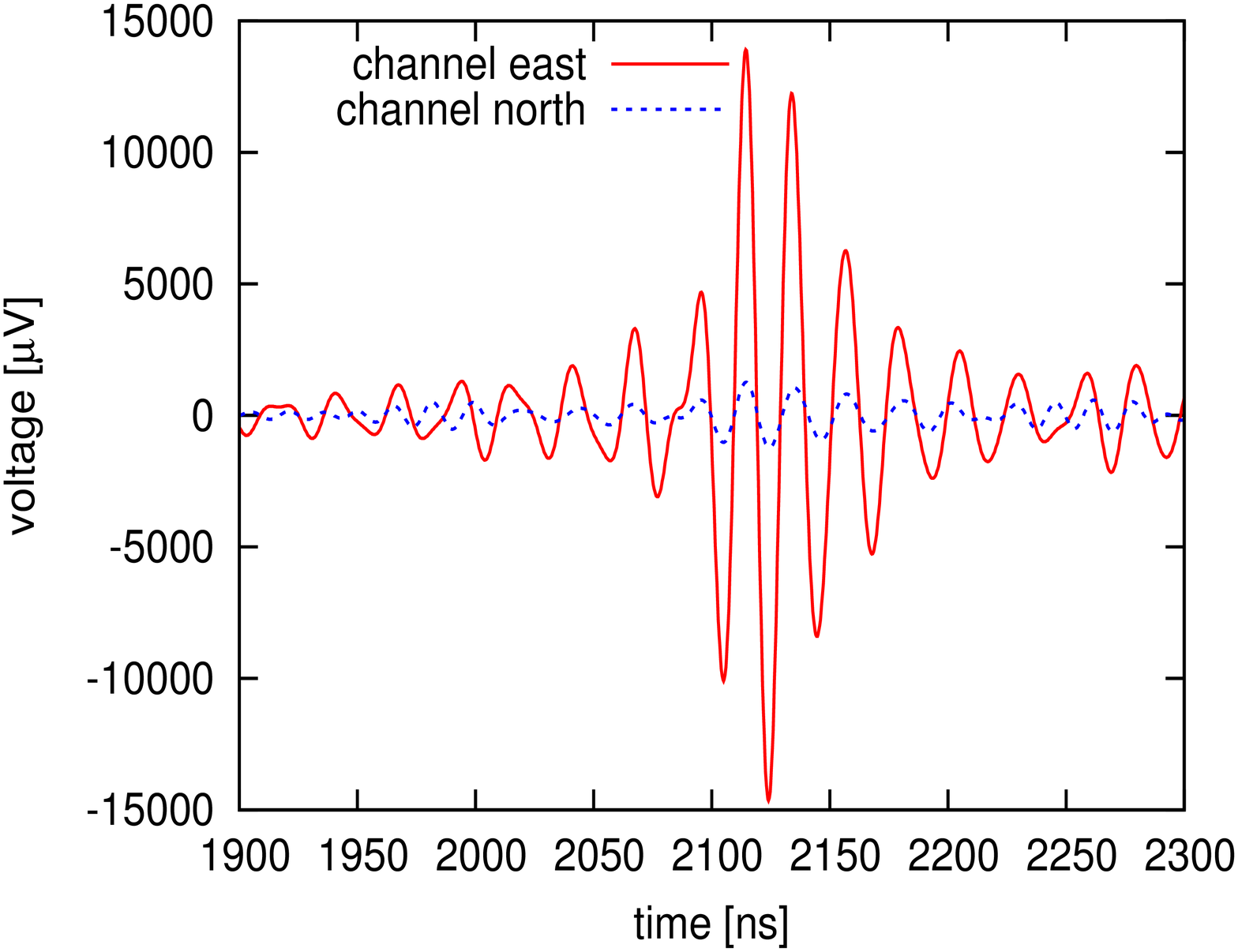,width = .85\linewidth}
\caption{
Time traces of a simulated radio signal in two channels of the AERA detector~\cite{radio}.
}\label{fig-mhz}
\end{minipage}
\end{figure}

The study of cosmic rays with energies between $10^{17}$ and $5\times10^{18}$\,eV are of special interest. The transition
from a galactic to an extragalactic flux might occur in this energy range. A discrimination between astrophysical models requires
a precise measurement of the spectrum as well as the chemical composition. Two extensions have been built in the Auger experiment.
High Elevation Auger Telescopes (HEAT) are three fluorescence telescopes which can be tilted 45$^{\circ}$ above the horizon. They
extend the lower energy threshold to well below $10^{17}$\,eV. A shower measured by a HEAT telescope and one standard fluorescence
telescope is shown in Fig.~\ref{fig-heat}. This is a low energy shower which could be reconstructed due to the additional measurement
at higher elevation.
There is also an extension of the surface detector close to the HEAT site which is going to be combined with underground muon counters.
This infill array lowers the energy threshold of the surface detector down to $\sim5\times10^{17}$\,eV.

Other activities are connected with the radio detection of EAS.
Radio detection might play an important role in the future, because a low-cost radio detector could have 100\% uptime and sensitivity
to the chemical composition.
Radio emission from EAS in the frequency range of 30--80\,MHz is detected
by a prototype radio telescope array (AERA -- Auger Engineering Radio Array)~\cite{aera}. A simulated radio signal is shown in Fig.~\ref{fig-mhz}.
Predicted microwave emission~\cite{gorham} is going to be investigated by parabolic antennas equipped with a matrix of receivers measuring
in C~band (3.4--4.2\,GHz) and Ku~band (10.95--14.50\,GHz).

\section*{Conclusions}

The Pierre Auger Observatory and its results have been described. The observatory provides valuable
data for the study of ultra-high energy cosmic rays. Major achievements of the observatory are the precise
measurement of the cosmic ray flux and chemical composition, as well as the detailed study of arrival directions.
In addition, the observatory has become the basis for testing several new detection methods.

\end{document}